\documentclass[twoside,leqno]{article}

\usepackage[letterpaper]{geometry}

\usepackage{ltexpprt}
\usepackage{hyperref}

\usepackage{amsmath}
\DeclareMathOperator{\erfc}{erfc}

\usepackage{listings}
\usepackage{color}

\usepackage{caption}

\usepackage[flushleft]{threeparttable}

\definecolor{dkgreen}{rgb}{0,0.6,0}
\definecolor{gray}{rgb}{0.5,0.5,0.5}
\definecolor{mauve}{rgb}{0.58,0,0.82}

\def\multiset#1#2{\ensuremath{\left(\kern-.3em\left(\genfrac{}{}{0pt}{}{#1}{#2}\right)\kern-.3em\right)}}

\DeclareRobustCommand{\bigO}{%
  \text{\usefont{OMS}{cmsy}{m}{n}O}%
}

\begin{document}

\newcommand\relatedversion{}
\renewcommand\relatedversion{\thanks{The full version of the paper can be accessed at \protect\url{https://arxiv.org/abs/2412.05300}}} 

\title{\Large AD-HOC: A C++ Expression Template package \protect\\ for high-order derivatives backpropagation\relatedversion}
\author{Juan Lucas Rey\thanks{Bloomberg}}

\date{}

\maketitle


\fancyfoot[R]{\scriptsize{Copyright \textcopyright\ 2025 by SIAM\\
Unauthorized reproduction of this article is prohibited}}





\begin{abstract} \small\baselineskip=9pt This document presents a new C++ Automatic Differentiation (AD) tool, AD-HOC (Automatic Differentiation for High-Order Calculations). This tool aims to have the following features:

  \begin{itemize}
    \item Calculation of user specified derivatives of arbitrary order
    \item To be able to run with similar speeds as handwritten code
    \item All derivatives calculations are computed in a single backpropagation tree pass
    \item No source code generation is used, relying heavily on the C++ compiler to statically build the computation tree before runtime
    \item A simple interface
    \item The ability to be used \textit{in conjunction} with other established, general-purpose dynamic AD tools
    \item Header-only library, with no external dependencies
    \item Open source, with a business-friendly license
  \end{itemize}
\end{abstract}

\section{Motivation}

AD has been successfully employed in finance across numerous institutions, offering significant computational benefits, especially in the wake of the 2006 financial crisis when complex risk calculations were imposed on banks. However, certain banks or fintech companies might not be willing to use AD techniques that use source code generation, mainly because of security concerns.

This paper introduces a new AD tool, AD-HOC (Automatic Differentiation for High-Order Calculations), which offers comparable run times to source code generation tools or handwritten code, while avoiding the need for source code generation. To achieve such performance, this tool employs novel C++17/20 techniques like expression templates, extensive use of template meta-programming, and the \verb|constexpr| keyword. Additionally, AD-HOC facilitates the calculation of high-order derivatives, utilising Taylor expansions to compute all derivatives up to any arbitrary order in a single backpropagation.

The technique for computing high-order derivatives within a single backpropagation pass is not novel and has been previously implemented in tools such as Arbogast \cite{CHARPENTIER} by Isabelle Charpentier et al. and COSY INFINITY \cite{COSY} by Martin Berz et al. Similarly, the utilisation of expression templates in AD tools is not new either, as it was previously done by M. Sagebaum et al. \cite{SAGEBAUM}, Eric Philips et al. \cite{PHIPPS}, Robin Hogan \cite{HOGAN}, Pierre Auber et al. \cite{AUBERT} and Antoine Savine \cite{SAVINE}. However, in this implementation, expression templates are employed differently, with types allowing discrimination between input variables at compile-time, containing zero data yet describing a complete calculation tree. The novelty of this approach relies on this new templetisation approach, allowing for faster runtimes, coupled with the combination of high-order Taylor backpropagation.

Another notable feature of AD-HOC is its ability to \textit{selectively} compute high-order derivatives required by the user (for example calculating $\frac{d^3(f(x_1, x_2))}{d(x_1)^2d(x_2)}$), \textit{avoiding} the computation of similarly defined high-order derivatives (for example $\frac{d^3(f(x_1, x_2))}{d(x_1)d(x_2)^2}$). This ensures optimal calculations tailored to the user's needs, particularly useful when pricing financial options using formulas like Black-Scholes, where practitioners typically require all first-order derivatives and a selection of second-order derivatives (such as Gamma, Vanna and Volga). AD-HOC enables the creation of a backpropagation tree that calculates only the necessary derivatives, minimising delays due to unnecessary computations.

Furthermore, AD-HOC has been used in conjunction with other, more general dynamic AD tools. Notably, dynamic AD tools such as dco/c++\footnote{\url{https://nag.com/automatic-differentiation/}}, CoDiPack\footnote{\url{https://github.com/SciCompKL/CoDiPack}}, and ADOL-C\footnote{\url{https://github.com/coin-or/ADOL-C}} allow the introduction of external functions that may be locally more efficient. While these external functions have typically handwritten derivatives, AD-HOC obviates the need for manual intervention and keeps the same runtime speed. Additionally, AD-HOC's ability to compute high-order derivatives enables it to provide external functions for second-order modes like adjoint-on-tangent.

\section{A brief algorithm description}

As mentioned before, the technique used in this algorithm, utilising high-order Taylor expansions, is not new. The sources of this technique can be traced back to Louis François Antoine Arbogast 1800's book `Du calcul des dérivations'. In particular, in the first article, section IV, part `Substitution des séries dans les séries', page 53., Arbogast provides a very detailed explanation of how to backpropagate derivatives with an example of 3 univariate functions\footnote{even though the term backpropagation would be coined some 150 years later! Arbogast's work is criminally under-appreciated}.

Let's assume we have a calculation tree described as a DAG (Directed Acyclic Graph) where elementary univariate ($\exp$, $\cos$,...), bivariate ($\times$, $+$, incomplete elliptic integral of the first kind,...) or even in some cases trivariate functions (incomplete elliptic integral of the third kind\footnote{these unfrequently used functions are now part of the C++17 standard}) are used in a particular order. We then execute our calculation following the natural or `forward' order of our DAG, keeping all intermediary results. Our task is now to calculate some arbitrarily defined high-order derivatives of our final result, and this is equivalent to finding a corresponding Taylor expansion, where the monomials present at the end of our algorithm match the required derivatives.

Let's take as an example the calculation of all second-order derivatives of $R = \exp(\cos(V1\times V2))$. We begin our backpropagation writing the Taylor expansion of the identity function applied on the result, introducing a perturbation on the variable $R$, namely $ \epsilon_{R}$: $I(R+\epsilon_{R}) = R+ \epsilon_{R}$. It might appear inconsequential to highlight such an obvious fact, but it is worth noting that this Taylor expansion is valid for any order, since $\frac{\partial^n I(x)}{\partial x^n} = 0$ for $n>1$. This means that the beginning of this algorithm, with a single seed per output, is valid for a Taylor expansion up to any order, and can therefore accommodate any number of high-order derivatives required by our algorithm.

We now introduce a variable $Q=\cos(V1*V2)$, and an associated perturbation $ \epsilon_{Q}$, for which we have the equation $R=\exp(Q)$. Given that we only need the second-order derivatives of the final result, we only need to expand the $\exp$ univariate function up to order 2. We can then write the truncated Taylor expansion:

\begin{equation*}
  \exp(Q + \epsilon_{Q}) = \exp(Q) + \frac{\exp'(Q)}{1!} \epsilon_{Q} + \frac{\exp''(Q)}{2!} \epsilon_{Q}^2 + \bigO( \epsilon_{Q}^3)
\end{equation*}

Assuming we choose an infinitesimal value of $\epsilon_{Q}$ such that we have $R+ \epsilon_{R} = \exp(Q + \epsilon_{Q})$, we infer that we have $\exp(Q) + \epsilon_{R} = \exp(Q + \epsilon_{Q})$ and therefore:

\begin{equation}
  \epsilon_{R} = \frac{\exp'(Q)}{1!} \epsilon_{Q} + \frac{\exp''(Q)}{2!} \epsilon_{Q}^2 + \bigO( \epsilon_{Q}^3)\label{eq:1}
\end{equation}

Similarly, introducing the variable $P=V1 \times V2$ and $\epsilon_{P}$, for which we then have $Q = \cos(P)$, we infer:

\begin{equation}
  \epsilon_{Q} = \frac{\cos'(P)}{1!} \epsilon_{P} + \frac{\cos''(P)}{2!} \epsilon_{P}^2 + \bigO( \epsilon_{P}^3)\label{eq:2}
\end{equation}

We can now replace all values of $\epsilon_{Q}$ in \eqref{eq:1} by the Taylor expansion in the series \eqref{eq:2} (doing what Arbogast called a `substitution of a series in a series'), to obtain:

\begin{equation}
  \begin{split}
    \epsilon_{R} &= \frac{\exp'(Q)}{1!} \left( \frac{\cos'(P)}{1!} \epsilon_{P} + \frac{\cos''(P)}{2!} \epsilon_{P}^2 + \bigO( \epsilon_{P}^3) \right) + \frac{\exp''(Q)}{2!}  \left( \frac{\cos'(P)}{1!} \epsilon_{P} + \frac{\cos''(P)}{2!} \epsilon_{P}^2 + \bigO( \epsilon_{P}^3) \right) ^2 + \bigO( \epsilon_{P}^3) \\
    &=  \alpha \epsilon_{P} +  \beta \epsilon_{P}^2 + \bigO( \epsilon_{P}^3) \text{  \ \  with  \ \  }  \alpha = \frac{\exp'(Q)\cos'(P)}{1!1!} \text{ \ \  and \ \  }  \beta = \frac{\exp'(Q)\cos''(P)}{1!2!}  + \frac{\exp''(Q)\cos'(P)^2}{2!1!^2}\label{eq:3}
  \end{split}
\end{equation}

Finally for $V1$ and $V2$ we have $  (V1+ \epsilon_{V1}) \times (V2+ \epsilon_{V2}) = V1 \times V2 + V2 \times \epsilon_{V1} + V1 \times \epsilon_{V2} + \epsilon_{V1} \times \epsilon_{V2}$.

Again, assuming we choose infinitesimal values of $\epsilon_{V1}$ and $\epsilon_{V2}$ such that $P+ \epsilon_{P} = (V1+ \epsilon_{V1}) \times (V2+ \epsilon_{V2})$, we have:

\begin{equation}
  \epsilon_{P}  = V2 \times \epsilon_{V1} + V1 \times \epsilon_{V2} + \epsilon_{V1} \times \epsilon_{V2}\label{eq:4}
\end{equation}

We can finally replace all values of $\epsilon_{P}$ in \eqref{eq:3} by the Taylor expansion in the series \eqref{eq:4} to obtain the final Taylor expansion:

\begin{equation*}
  \begin{split}
    \epsilon_{R}  &= \alpha \left(V2 \times \epsilon_{V1} + V1 \times \epsilon_{V2} + \epsilon_{V1} \times \epsilon_{V2} \right) + \beta \left(V2 \times \epsilon_{V1} + V1 \times \epsilon_{V2} + \epsilon_{V1} \times \epsilon_{V2} \right)^2 + \bigO( \epsilon_{Vi}^3) \\
    &= \alpha V2 \epsilon_{V1} + \alpha V1 \epsilon_{V2} + (\alpha + 2 \beta V1 V2)\epsilon_{V1} \times \epsilon_{V2} + \beta V2^2 \epsilon_{V1}^2 + \beta V1^2 \epsilon_{V2}^2 + \bigO( \epsilon_{Vi}^3) \\
    \bigO( \epsilon_{Vi}^3) &= \bigO( \epsilon_{V1}^3) + \bigO( \epsilon_{V1}^2\epsilon_{V2}) + \bigO( \epsilon_{V1}\epsilon_{V2}^2) + \bigO( \epsilon_{V2}^3)
  \end{split}
\end{equation*}

Once this backpropagation is performed, lastly we multiply the remaining Taylor coefficients by the corresponding factorials to be able to obtain the desired derivatives.

This example shows a second-order backpropagation, but could have been performed with a higher order, by keeping larger expansions at every stage instead of truncating at order 2. Also, the subtleties of truncating certain intermediate monomials whenever they are not needed, has not been described, but has been implemented in the algorithm.

\section{AD-HOC C++ interface description}

AD-HOC requires C++17, though C++20 is the preferred choice, even if this only improves the interface. Even though expression templates have been used before, to the best of the author's knowledge, the following used techniques are new, namely:

\begin{itemize}
  \item Every input has a unique type, all different from each other
  \item Constants have their own type, capable of describing up to a 64-bit float number without occupying any allocated memory at runtime
  \item All memory allocation is in the stack exclusively using \verb|std::array|
  \item The (forward) primal calculation tree and the (backward) backpropagator are separate classes. All the allocated static memory used by the algorithms is contained in those 2 classes
  \item This means that our input, intermediary or final variables contain/own no allocated data, not even a single float nor a reference to a float. They are just descriptive types
\end{itemize}

In the following interface description, the Black-Scholes formula \cite{BLACK} will be used as an example to evaluate the price of a financial option. As a reminder:

\begin{equation*}
  C(S, K, \sigma, T, r) = N(d_1)S - N(d_2)Ke^{-rT}  \text{  with  } d_1 = \frac{\ln \frac{S}{K} + (r+\frac{\sigma^2}{2})T}{\sigma \sqrt T} \text{  and  }   d_2 = d_1 - \sigma \sqrt T
\end{equation*}

With $S$ representing the spot value, $K$ the option strike, $\sigma$ the volatility, $T$ the time to maturity, and $r$ the instantaneous interest rates. These 5 variables and their associated types will be represented in the following code respectively by \verb|S|, \verb|K|, \verb|V|, \verb|T|, \verb|R|.

\subsection{Input Variables.}

The input variables of the calculation are each defined by distinct types. To facilitate this, we use a C++20 feature where a string can be used as a template parameter. We can then define the $S$ spot variable with a simple statement: \verb|adhoc::double_t<"S"> S;| Thanks to a helpful macro \verb|#define ADHOC(x) double_t<#x> x|, we can define a variable in an even simpler way: \verb|ADHOC(S);|. This string now not only serves as a compile-time type identifier but also serves to print meaningful information when a static assert triggers a compiler error. Using this macro also ensures that the uniqueness of a variable declaration on a C++ program also forces the uniqueness of its associated type.

\subsection{Constant Types.}

AD-HOC aims to fully depict the calculation tree within a single data-less type. Consequently, even a constant float like $\pi$ or any arbitrary float must be accommodated within this type, as exemplified in a function such as $f(x) = x \times \pi$. In C++20 this task is straightforward since a float type can serve as a template parameter. For this, a class is needed (here the Constant Double class \verb|CD|) that can wrap this constant in a type. Using this class, a simple cumulative distribution function (CDF) for the normal distribution can be written as follows:

  \begin{lstlisting}
  template <class D> inline auto cdf_n(D x) {
      constexpr double m_one_over_sqrt2 = -0.70710678118654757;
      return CD<0.5>() * erfc(x * CD<m_one_over_sqrt2>()); }
  \end{lstlisting}

Ideally, for an even more elegant syntax, a current C++ proposal allowing \verb|constexpr| arguments in a function would need to be adopted. This proposal, championed by David Stone \cite{AUBERT}, would allow the usage of an \verb|operator*| defined as follows:

   \begin{lstlisting}
  auto operator*(constexpr double rhs) const { return mul_t<Base, CD<rhs>()>{}; }
  \end{lstlisting}

In this case, the double constant can be used directly and the syntax would be then simplified to:

  \begin{lstlisting}
  template <class D> inline auto cdf_n(D x) {
      constexpr double m_one_over_sqrt2 = -0.70710678118654757;
      return 0.5 * erfc(x * m_one_over_sqrt2);}
  \end{lstlisting}

This function \verb|cdf_n| will return a variable with no data, of the following type:

  \begin{lstlisting}
  mul_t< C<double, Arg<double>{5.0e-1}>,
          erfc_t< mul_t< double_t<0>,
                           C<double, Arg<double>{-7.0710678118654757e-1}> > > >
  \end{lstlisting}

\subsection{Calculation Tree}

The full Black-Scholes calculation can be now written in C++20 as follows:

  \begin{lstlisting}
  template <class D> inline auto cdf_n(D x) {
      constexpr double m_one_over_sqrt2 = -0.70710678118654757;
      return CD<0.5>() * erfc(x * CD<m_one_over_sqrt2>()); }

  template <class I1, class I2, class I3, class I4, class I5> auto
  call_price(I1 S, I2 K, I3 V, I4 T, I5 R) {
      auto tvol = V * sqrt(T);
      auto d1 = log(S / K) / tvol + tvol * CD<0.5>();
      auto d2 = d1 - tvol;
      return S * cdf_n(d1) - K * cdf_n(d2) * exp(-R * T); }
  \end{lstlisting}

For the usage of the AD-HOC tool, the \verb|call_price| function would need to be instantiated with template parameters \verb|I1=double_t<"S">|, \verb|I2=double_t<"K">|, \verb|I3=double_t<"V">|, \verb|I4=double_t<"T">| and \verb|I5=double_t<"R">|. However, in the case where we don't use AD-HOC, we can simply use \verb|call_price| by instantiating with \verb|I1=I2=I3=I4=I5=double|.

As is customary with adjoint automatic differentiation, most intermediate values must be stored in a calculation tree, or `tape'. The reduction of this tape is paramount to obtain a fast program runtime \cite{NAUMANN}. The advantage of a static approach is that it allows us to not only analyse how many intermediate variables need to be allocated, but also how nodes will be used, thereby avoiding the storage of certain values. For example, consider the final operator of Black-Scholes, where we have a subtraction: $N(d_1)S - N(d_2)Ke^{-rT}$. In this case, we do not need to store the values of $N(d_1)S$ and $N(d_2)Ke^{-rT}$ in our calculation tree, as the backpropagation algorithm does not require these values to provide the corresponding derivative values of 1 and -1. Additionally, some operators, like $\erfc(x)$, only need the input value $x$ to calculate derivatives of order 1 or higher, while others, like $\tan(x)$ or $\exp(x)$, only need the output value. Consequently, if we had a calculation involving $\tan(\erfc(x))$, we know that the value of $\erfc(x)$ does not need to be stored in the calculation tree for backpropagation (assuming that node is not used elsewhere), as it is unnecessary for the backpropagation of either the $\tan$ or $\erfc$ operators. These analyses and memory optimisations can also be performed with a dynamic automatic differentiation tool, but the cost is incurred at runtime and may not be beneficial overall. This is because with a dynamic tool, we cannot infer any information about how a particular variable is used at compile-time. However, with a static tool, the benefits in terms of memory allocation for these analyses are undeniable, as the cost is incurred only at compile-time.

The Calculation Tree creation and usage would look like this in C++20:

  \begin{lstlisting}
  ADHOC(S); ADHOC(K); ADHOC(V); ADHOC(T); ADHOC(R);
  auto Price = call_price(S, K, V, T, R);                      // This line just creates an empty type
  adhoc::CalcTree ct(Price);                                   // Allocation of the necessary tape memory
  ct.set(S) = 100.0; ct.set(K) = 102.0; ct.set(V) = 0.15; ct.set(T) = 0.5; ct.set(R) = 0.01;
  ct.evaluate(); std::cout << ct.get(Price) << std::endl;  // Calculations are actually done here
  \end{lstlisting}

At this juncture, it's worth noting that the variable \verb|Price| is once again empty, but its type is sufficient to describe the entire Black-Scholes calculation, including constants. It's only during the instantiation of the type \verb|CalcTree| that we allocate all the memory required to store the necessary intermediate variables, in a single, stack-allocated \verb|std::array|.

\subsection{Differential Operator}
Since AD-HOC will require the user to identify high-order derivatives to calculate, the operator \verb|adhoc::d| is introduced, for both input and output derivatives. For example, considering a function $f(x_1, x_2)$ for which we require the derivative $\frac{\partial^2 f}{\partial x_1 \partial x_2}$, this will be represented by \verb|d(x_1)*d(x_2)| in the code. Similarly, we will use \verb|d<2>(x_1)| to represent $\frac{\partial^2 f}{(\partial x_1)^2}$. The numerator $\partial^2 f$ is implied in both cases and should correspond to what's at the top level of the backpropagation tree.

Assuming we have a calculation tree with multiple outputs, namely $f_1(x_1, x_2)$ and $f_2(x_1, x_2)$, both with their corresponding user-provided seeds $s_1$ and $s_2$, then \verb|d(x_1)*d(x_2)| would represent the value:

\begin{equation*}
  \frac{\partial^2 (s_1 \times f_1 + s_2 \times f_2)}{\partial x_1 \partial x_2} = s_1 \times \frac{\partial^2 ( f_1 )}{\partial x_1 \partial x_2} + s_2 \times \frac{\partial^2 (f_2)}{\partial x_1 \partial x_2}
\end{equation*}

\subsection{Backpropagator}\label{Backpropagator}
This section shows how the backpropagation tree is defined and used. For first-order derivatives, creating such a tree is relatively straightforward as it is a mirror of the primal calculation tree. However, as derivatives of order 2 or higher are requested, the backpropagation tree will diverge and require additional nodes.

AD-HOC makes sure that throughout the backpropagation process, only the current slice needs to be retained in static memory until we complete the backpropagation. Once again, leveraging the static nature of our tool, we are able to analyse the maximum memory size utilised during backpropagation at compile-time, ensuring that only an array of that size is allocated, henceforth referred to as a buffer. As mentioned earlier, a dynamic automatic differentiation tool could also conduct such an analysis, but it would entail a runtime cost with unclear benefits, facing the customary trade-off between time and space. Here is a usage example of the BackPropagator class:

  \begin{lstlisting}
  using adhoc::d;
  auto bp = BackPropagator( /* input */ d(S), d(V), d<2>(S), /* output */ d(Price));
  bp.set(d(Price)) = 1.; bp.backpropagate(ct);
  std::cout << bp.get(d(S)) << ", " << bp.get(d(V)) << ", " << bp.get(d<2>(S)) << std::endl;
  \end{lstlisting}

It's noteworthy that the arguments of the BackPropagator can be arranged in any order, and multiple output derivatives can even be included as arguments to it. In this specific scenario, the variable `\verb|Price|' constitutes the top level of the entire calculation tree. Consequently, the seed for the backpropagation process is solely \verb|d(Price)|.

Also, at this stage we can see that the differential operator \verb|d(K)| is not included in the \verb|BackPropagator|. This is because the strike of an option is a constant and does not evolve with market conditions, therefore no derivative to that parameter is required by practitioners. When we omit this differential operator, any calculation regarding the derivative of that parameter will also be omitted, that parameter will be considered then a `passive' one.

Additionally, the calculation tree instance \verb|ct| is passed as an argument only during the backpropagation phase. This facilitates a single \verb|BackPropagator| instance capable of backpropagating multiple \verb|CalculationTree| instances, with potentially varying inputs and even types. There are multiple reasons to separate the \verb|CalculationTree| from the \verb|BackPropagator|. For a PDE grid, we might need the same type of evolution repeated for every time-step, and all those would have to be stored in multiple \verb|CalculationTree| instances, but only one \verb|BackPropagator| (and associated \verb|std::array| buffer) is needed. Also, AD-HOC has been used as an external function in conjunction with other tools such as dco/c++, CoDiPack and ADOL-C, and being able to separate the \verb|CalculationTree| from the \verb|BackPropagator| has proved useful in this scenario: the external function only stores what's in the \verb|CalculationTree| instance, while calling the \verb|BackPropagator| only when requested during the backpropagation.

At the top level of the backpropagation, only the argument \verb|d(GlobalOutput)| (or \verb|d(GlobalOutput_1)|, \verb|d(GlobalOutput_2)|,... if there are multiple outputs) is necessary for our \verb|BackPropagator|. However, we could also use known AD techniques like `checkpointing' by having multiple subsequent calculation trees, and calling their respective backpropagators in inverse order. Given that we are dealing with high-order derivatives with AD-HOC, particular care would have to be taken to backpropagate all the order of derivatives during the checkpointing.

\section{Application example: Vanna-Volga method}

It has become market practice in finance to calibrate implied volatility curves and approximately asses risk using a method called Vanna-Volga. This method, popularised by Mercurio et al. \cite{MERCURIO} requires the calculation of Vega, ($\frac {\partial C}{\partial \sigma}$), Vanna ($\frac {\partial^2 C}{\partial \sigma \partial S}$) and Volga ($\frac {\partial^2 C}{(\partial \sigma)^2}$) of multiple Black-Scholes options in the market. Given that these calculations are used in the calibration of local vol surfaces, with repeated executions, the majority of the runtime would be spent doing these, therefore any performance improvement is desirable. We will therefore benchmark the runtime of a simple Black-Scholes price against the calculation of the same price and these 3 additional \textit{cherry-picked} derivatives. The code for calculating Black-Scholes can be taken as defined in the previous section and instantiated with a double type as follows (this is referred as the BASE version of the code):

  \begin{lstlisting}
  double S = 100., K = 102., V = 0.15, T = 0.5, R = 0.01;
  double Price = call_price(S, K, V, T, R); std::cout << Price << std::endl;
  \end{lstlisting}

For the calculation of the Price and our 3 derivatives using AD-HOC, the code looks as follows (AD-HOCv1):

  \begin{lstlisting}
  ADHOC(S); ADHOC(K); ADHOC(V); ADHOC(T); ADHOC(R);
  auto Price = call_price(S, K, V, T, R);
  CalcTree ct(Price);
  ct.set(S) = 100.0; ct.set(K) = 102.0; ct.set(V) = 0.15; ct.set(T) = 0.5; ct.set(R) = 0.01;
  ct.evaluate();
  std::cout << ct.get(Price) << std::endl;

  auto bp = BackPropagator( /* input */ d(V), d<2>(V), d(V) * d(S), /* output */ d(Price));
  bp.set(d(Price)) = 1.;
  bp.backpropagate(ct);
  std::cout << bp.get(d(V)) << ", "<< bp.get(d<2>(V)) << ", " << bp.get(d(V) * d(S)) << std::endl;
  \end{lstlisting}

This version only increases the runtime by close to 45\% to obtain 3 additional greeks, 2 of which are of second-order.

The following code is added for a first version to calculate these greeks by hand (version refered to as HANDv1):

  \begin{lstlisting}
  inline auto pdf(double x) { return 0.39894228040143265 * exp(-0.5 * x * x); }

  inline auto vega(double S, double K, double V, double T) {
      auto tvol = V * std::sqrt(T);
      auto d1 = std::log(S / K) / tvol + tvol * 0.5;
      return S * pdf(d1) * std::sqrt(T); }

  inline auto vanna(double S, double K, double V, double T) {
      auto tvol = V * std::sqrt(T);
      auto d1 = std::log(S / K) / tvol + tvol * 0.5;
      auto d2 = d1 - tvol;
      return -pdf(d1) * d2 / v; }

  inline auto volga(double S, double K, double V, double T) {
      auto tvol = V * std::sqrt(T);
      auto d1 = std::log(S / K) / tvol + tvol * 0.5;
      auto d2 = d1 - tvol;
      return S * pdf(d1) * d1 * d2 * T / tvol; }
  \end{lstlisting}

With these hand-written functions we get a version that is around 20\% slower than what AD-HOCv1 can achieve. To truly give handwritten code a chance we will have to write an ad-hoc (pardon the pun) function implementation where the Price, Vega, Vanna and Volga are calculated in a single function (HANDv2):

  \begin{lstlisting}
  inline auto call_price_vega_vanna_volga(double S, double K, double V, double T, double R) {
      std::array<double, 4> results;
      auto sqrtT = std::sqrt(T);
      auto tvol = V * sqrtT;
      auto d1 = std::log(S / K) / tvol + tvol * 0.5;
      auto d2 = d1 - tvol;
      auto pdfd1 = pdf(d1);
      results[0] = S * cdf_n(d1) - K * cdf_n(d2) * exp(-R * T);   // price
      results[1] = S * pdfd1 * sqrtT;                                 // vega
      results[2] = -pdfd1 * d2 / V;                                   // vanna
      results[3] = S * pdfd1 * d1 * d2 * T / tvol;                    // volga
      return results; }
  \end{lstlisting}

This 2nd version is faster than AD-HOCv1 by around 20\%, due to the fact that the handwritten Vega formula used is a \textit{symbolic} simplification of a straightforward derivative calculation:

\begin{alignat*}{2}
  \frac{\partial C}{ \partial \sigma} &= \frac{e^{-\frac{d_1^2}{2}}}{\sqrt{2\pi}}  \frac{\partial d_1}{ \partial \sigma} S -  \frac{e^{-\frac{d_2^2}{2}}}{\sqrt{2\pi}}  \frac{\partial d_2}{ \partial \sigma} K e^{-rT}
  && = \frac{e^{-\frac{d_1^2}{2}}}{\sqrt{2\pi}}  \left(\frac{\partial d_1}{ \partial \sigma} S -  e^{\frac{d_1^2}{2} - \frac{d_2^2}{2}}  \frac{\partial d_2}{ \partial \sigma} K e^{-rT} \right) \\
  & = N'(d_1)  \left(\frac{\partial d_1}{ \partial \sigma} S -  e^{\frac{1}{2} (d_1+d_2)(d_1-d_2)}  \frac{\partial d_2}{ \partial \sigma} K e^{-rT} \right)
  && = N'(d_1)  \left(\frac{\partial d_1}{ \partial \sigma} S -  e^{\frac{1}{2} \frac{2 \ln\frac{S}{K} + 2rT}{\sigma \sqrt{T}}\sigma \sqrt{T}}  \frac{\partial d_2}{ \partial \sigma} K e^{-rT} \right) \\
  & = N'(d_1)  \left(\frac{\partial d_1}{ \partial \sigma} S -   \frac{\partial d_2}{ \partial \sigma} S \right)
  && = N'(d_1)  S \left(\frac{\partial d_1}{ \partial \sigma} -   \frac{\partial (d_1 - \sigma \sqrt{T}) }{ \partial \sigma} \right) \\
  & = N'(d_1)  S \left(\frac{\partial d_1}{ \partial \sigma} -   \frac{\partial d_1  }{ \partial \sigma}  + \sqrt{T} \right)
  && = N'(d_1)  S \sqrt{T}
\end{alignat*}

As it is often claimed, Automatic Differentiation is not Symbolic Differentiation, and an AD tool does not aspire to perform such simplifications. We could say that this simplification gives both hand-written code versions an `unfair' advantage, and it's unusual to find a situation where a derivative can be simplified symbolically at this level. Still, with this knowledge, we can still use AD-HOC to incorporate this new information and provide a similar performance. We can write both the Price and the (symbolically simplified) Vega as types:

  \begin{lstlisting}
  template<class D> inline auto pdf_n(D x) {
      constexpr double one_over_root_two = 0.39894228040143265;
      return CD<one_over_root_two>() * exp(CD<-0.5>() * x * x); }

  template<class I1, class I2, class I3, class I4>
  inline auto vega(I1 S, I2 K, I3 V, I4 T) {
      auto tvol = V * sqrt(T);
      auto d1 = log(S / K) / tvol + tvol * CD<0.5>();
      return S * pdf_n(d1) * sqrt(T); }

  ADHOC(S); ADHOC(K); ADHOC(V); ADHOC(T); ADHOC(R);
  auto Price = call_price(S, K, V, T, R);
  auto Vega = vega(S, K, V, T);

  CalcTree ct(Price, Vega);            // Price and Vega are both part of the calculation tree now
  ct.set(S) = 100.0; ct.set(K) = 102.0; ct.set(V) = 0.15; ct.set(T) = 0.5; ct.set(R) = 0.01;
  ct.evaluate(); std::cout << ct.get(Price) << ", " << ct.get(Vega) << std::endl;
  \end{lstlisting}

Our backpropagator now can just calculate Vanna and Volga as derivations of the simplified Vega expression:

  \begin{lstlisting}
  auto bp = BackPropagator( /* input */ d(Volatility), d(Spot), /* output */ d(Vega));
  bp.set(d(Vega)) = 1.; bp.backpropagate(ct);
  std::cout << bp.get(d(Volatility)) << ", " << bp.get(d(Spot)) << std::endl;
  \end{lstlisting}

With this new AD-HOCv2 version, the runtimes are almost identical to the ones from HANDv2 and AD-HOC has been able to match some very specific tailored handwritten code, even incorporating information from a symbolic simplification. Following are the specific timing measurements\footnote{ran on Apple M3 Pro, GCC compiler 13.2.0, using -O3 optimisations, and multiple repetitions with randomized inputs.}:

\begin{center}
  \begin{tabular}{||c c c c c c c c c c||}
    \hline
    Reps. & BASE & AD-HOCv1& R & HANDv1 & R & AD-HOCv2 & R & HANDv2 & R \\ [0.5ex]
    \hline\hline
    1M & 37$\mu$s & 45$\mu$s & 1.21x & 60$\mu$s & 1.62x & 44$\mu$s & 1.19x & 46$\mu$s & 1.24x \\
    \hline
    10M & 307$\mu$s & 431$\mu$s & 1.40x & 503$\mu$s & 1.64x & 350$\mu$s & 1.14x & 358$\mu$s & 1.17x \\
    \hline
    100M & 2900$\mu$s & 4224$\mu$s & 1.45x & 4992$\mu$s & 1.72x & 3487$\mu$s & 1.20x & 3600$\mu$s & 1.24x \\
    \hline
  \end{tabular}
\end{center}

As a conclusion, AD-HOC can provide similar performance to highly specific hand-written code, and the user can obtain 3 additional greeks only for an additional 20\% computation time. From a practitioner's point of view, it is not practical to have to handwrite a separate, specialised implementation of Black-Scholes adding specific greeks for each separate utilisation. Not only will this lead to cluttered, unmaintainable code with many specific micro-optimisations, but it is also prone to error, especially for high-order derivatives that might not have well established known formulas as these ones. For this, AD-HOC can help creating competitively optimised code for each separate requirement, with a quick development cycle, and it can even accommodate with ease scenarios where symbolic simplifications are to be leveraged.

\section{Runtime comparison}

A financial valuation service provider, having to reprice large portfolios for its clients and aiming to give updated valuations throughout the day, may choose to calculate an expensive Taylor expansion of these portfolios overnight, and later update the prices of these portfolios as market conditions evolve through the day using this pre-calculated, stored Taylor expansion. In that context, a higher Taylor expansion can increase precision during these intraday `instantaneous' reports. Hence, we have an incentive to make this Taylor expansion as high as possible, given that these daily updates are the result of a simple polynomial calculation and are relatively inexpensive. Given that most portfolios contain a large quantity of options priced with the Black-Scholes formula, multiple runtimes will be compared with full tensors up to order 5, using multiple AD known packages. The code used to perform these timings is publicly available.\footnote{\url{https://github.com/juanlucasrey/AD-HOC/tree/main/case_studies/2024ADChicago}}

For reference, a formula with $n$ inputs has $\multiset{n}{k}$ distinct derivatives of order $k$, $\multiset{n}{k}$ being the multiset defined as $\multiset{n}{k} = \binom {n+k-1}{k}$. This means that a full tensor up to order $d$ will have to produce $\sum_{k=0}^{d} \multiset{n}{k} =\multiset{n+1}{d}= \binom {n+d}{d}$ numbers (note that the first term $\multiset{n}{0} = 1$ corresponds to the actual primal function evaluation). This means that, for our case where $n=4$, for orders 0 to 5 we have to produce 1, 5, 15, 35, 70 and 126 values respectively. The following table shows the runtime ratios $R(N)=\frac{\text{runtime}(N^{th}\text{ order Tensor})}{\text{runtime}(\text{function evaluation})}$, and $RR(N)=\frac{\text{runtime}(N^{th}\text{ order Tensor})}{\text{runtime}(N-1^{th}\text{ order Tensor})}$. These metrics are displayed for eight AD packages, namely AD-HOC, ADOL-C, dco/c++\footnote{dco/c++ and CoDiPack have vectorisation features, and dco/map has CUDA features that could not be leveraged, due to the fact that we are using Apple's M3 processor \label{refnote}}, dco/map\footref{refnote}, dco/codegen, CoDiPack\footref{refnote}, Enzyme and Tapenade.

\begin{center}
  \begin{tabular}{||c | c c c c||}
    \hline
    tool& $order$ & $R(order)$ & $RR(order)$ & outputs \\ \hline  \hline
    primal & 0 & 1x & -& 1 \\ \hline
    AD-HOC & 1 & 1.27x & 1.27x & 5\\
    AD-HOC & 2 & 1.66x & 1.30x & 15 \\
    AD-HOC & 3 & 2.60x & 1.56x &35 \\
    AD-HOC & 4 & 6.96x & 2.67x & 70 \\
    AD-HOC & 5 & 16.66x & 2.39x & 126 \\ \hline
    ADOL-C & 1 & 23.0x & 23.0x & 5 \\
    ADOL-C & 2 & 57.68x & 2.50x & 15 \\
    ADOL-C & 3 & 139.0x & 2.41x & 35 \\
    ADOL-C & 4 & 300.7x & 2.16x & 70 \\
    ADOL-C & 5 & 634.1x & 2.10x & 126 \\
    ADOL-C & 6 & 1142.73x & 1.80x & 210 \\
    ADOL-C & 7 & 2056.78x & 1.79x & 330 \\ \hline
    ADOL-C fwd & 1 & 8.36x & 8.36x & 5 \\ \hline
    dco/codegen & 1 & 1.51x & 1.51x & 5 \\ \hline
    dco/map & 1 & 1.92x & 1.92x & 5 \\ \hline
    dco/c++ & 1 & 3.03x & 3.03x & 5 \\
    dco/c++ & 2 & 13.93x & 4.60x & 15 \\ \hline
    CoDiPack & 1 & 3.63x & 3.63x & 5 \\
    CoDiPack & 2 & 16.36x & 4.50x & 15 \\ \hline
    Tapenade & 1 & 2.03x & 2.03x & 5 \\
    Tapenade & 2 & 6.66x & 3.28x & 15 \\ \hline
    Enzyme & 1 & 1.53x & 1.53x & 5 \\ \hline
  \end{tabular}
\end{center}

A number of factors explain these runtime differences. A first source of differences is that ADOL-C, dco/c++ and CoDiPack are \textit{dynamic} tools and can work on a much larger number of diverse scenarios, however their speed is usually slower. On the other hand, dco/codegen, dco/map, Enzyme, Tapenade and AD-HOC are \textit{static} tools and this allows for better runtime performances. This explains the relatively high first-order derivative ratio $R(1)$ for \textit{dynamic} tools (ADOL-C: 8.36x, dco/c++: 3.03x and CoDiPack: 3.63x) as opposed to \textit{static} tools (AD-HOC: 1.27x, dco/codegen: 1.51x, dco/map: 1.92x, Tapenade: 2.03x, Enzyme: 1.53x).

A second source of runtime differences, evidenced for second-order calculations and beyond, is that dco/c++, CoDiPack and Tapenade have to resort to backward-over-forward(-over-forward...) methods for high-order calculations. On the other hand, ADOL-C and AD-HOC perform a more efficient full Taylor backpropagation. This explains the relatively high second-order derivative $RR(2)$ for backward-over-forward methods (dco/c++: 4.60x, CoDiPack: 4.50x, Tapenade: 3.28x) as opposed to Taylor backpropagation methods (AD-HOC: 1.18x, ADOL-C: 2.5x). Furthermore, $RR(order)$ values remain between 2.0x and 2.5x and decrease for both AD-HOC and ADOL-C when calculating orders 3 and beyond. These are expected to remain at 4 (the number of inputs) for backward over iterative forward methods.

Theoretically, for Taylor backpropagation methods, $R(order)$ should be quadratic in the $order$ parameter \cite{GRIEWANK} and this should be seen reflected in a downward trend for $RR(order)$. ADOL-C clearly shows this trend, however, AD-HOC, does not show it as clearly, at least for the first 5 orders. This could be due to a variety of technical reasons, like cache misses due to the memory growing up to a point where processors lose certain efficiencies. Higher orders should be obtained to see further into this trend. At the time of writing, AD-HOC is being optimised further to be able to reach higher orders, the main difficulty being compilation times. A special mention is made, finally, to ADOL-C that can reach much higher orders than the ones performed in this exercise. A tool performing backward-over-iterative forward is expected to have a runtime ratio close to $R(order) = R(1) \times n^{order-1}$ and that value would be of at least 4096x for $n=4$ and $order=7$, where ADOL-C has a ratio of around 2000x. This clearly shows the superiority of the Taylor backpropagation algorithm for these types of calculations.

\section{Future directions}

AD-HOC is in the process of having features being added to expand its use. In particular, branching and high-order functions are being developed. For high-order functions, `reduce', `map', and `fold' like functions could be implemented using the new pipe operator `$|$', introduced in C++23\footnote{\url{https://en.cppreference.com/w/cpp/ranges/range\_adaptor\_closure}}, as to mimic a `C++ standard library' style:

\begin{lstlisting}
// types provided by AD-HOC
template <class T, class UnaryOperation> class RangeAdaptorClosure_t {};
template <typename T> class view {
  public:
    template <class UnaryOperation> auto operator|(UnaryOperation op) const {
        return RangeAdaptorClosure_t<view<T>, decltype(op(T{}))>{}; } };
// user code
view<double_t<"something">> data;
auto lambda = []<typename T>(T const &lhs) { return lhs * lhs; };
auto squared_data = data | lambda;
\end{lstlisting}
The variable \verb|squared_data| now contains all the necessary information to recreate the algorithm in both forward and backpropagation phase, while containing no concrete data:

\begin{lstlisting}
RangeAdaptorClosure_t< view < double_t<"something">  >,
                          mul_t< double_t<"something">, double_t<"something">  >  >
\end{lstlisting}

Another interesting, unrelated direction could be the inclusion of a user-defined compile-time node elimination order, in order to improve performance in certain cases. In particular it would be interesting to see if the optimal node elimination order remains optimal for high order derivatives or if some adjustments would have to be made discriminating between $+$ operators (of linear $order$ complexity for elimination) as opposed to univariate and $\times$ operators (of quadratic $order^2$ complexity for elimination). This subject is part of a much broader, NP-Complete \cite{NAUMANN2} problem and is the focus of much research in AD \cite{GRIEWANK}.

\section{Conclusion}

Even though AD has been used extensively in finance already, the techniques pioneered by tools such as ADOL-C, Arbogast or COSY INFINITY, where high-order Taylor expansions are used to backpropagate derivatives, have not. With AD-HOC, a powerful enough tool to create high-order derivatives backpropagation trees, we can calculate second-order derivatives, very much needed in finance, or even higher-order derivatives. In particular this tool could be useful for portfolio risk calculations, usually power hungry and requiring overnight calculations, or also in high-frequency trading scenarios, always requiring faster execution runtimes, and better precision for market-making requirements. Typically with these algorithms, the runtime increases in polynomial time \cite{BERZ} with the order of derivatives requested, and this is far superior than the exponential time required when using adjoint only for first-order derivatives, leaving the rest to iterations of tangent mode. AD-HOC package, open source and freely available\footnote{AD-HOC was started as a personal open source project: \url{https://github.com/juanlucasrey/AD-HOC}}, aims to at least familiarise practitioners to this fact with a tool that can claim some very competitive runtimes under certain conditions.

\end{document}